\documentclass[letterpaper]{article}

\usepackage[T1]{fontenc}

\usepackage{geometry}
\usepackage{setspace}

\usepackage{placeins}

\usepackage[
  backend=biber,
  style=numeric-comp,
  sorting=none,
  giveninits=true
]{biblatex}
\usepackage{booktabs}
\usepackage{graphicx}
\usepackage{float}
\newfloat{scheme}{htbp}{los}
\floatname{scheme}{Scheme}
\floatname{chart}{Chart}
\newfloat{graph}{htbp}{loh}
\usepackage[acronym]{glossaries}
\makeglossaries

\usepackage{xr-hyper}
\usepackage{hyperref}

\usepackage{chemformula} 
\usepackage[version = 4]{mhchem} 
\usepackage{siunitx}

\usepackage{etoc}

\usepackage{caption}
\DeclareCaptionLabelFormat{suppl}{#1~S#2}

\usepackage{authblk}
\author[1]{Anna Werkovits}
\author[1]{Simon B. Hollweger}
\author[1]{Oliver T. Hofmann*}
\affil[1]{Institute of Solid State Physics, Graz University of Technology, 8010 Graz, Austria}

\title{Emergent Rate Laws for Collective Lying–Standing Transitions}
\date{*Email: o.hofmann@tugraz.at}

\newacronym{kmc}{kMC}{kinetic Monte Carlo}
\newacronym{2sa}{IPL2SA}{Irreversible Power-Law Two-State Approximation}
\newacronym{dft}{DFT}{Density Functional Theory}
\newacronym{tst}{TST}{Transition State Theory}

\newcommand{\dELL}{\Delta E_{\mathrm{LL}}}
\newcommand{\dESS}{\Delta E_{\mathrm{SS}}}
\newcommand{\dESL}{\Delta E_{\mathrm{SL}}}
\newcommand{\dELS}{\Delta E_{\mathrm{LS}}}
\newcommand{\EadsL}{E_{\mathrm{ads,L}}}
\newcommand{\EadsS}{E_{\mathrm{ads,S}}}

\newcommand{\kLScol}{k_{\mathrm{LS},\mathrm{coll}}}
\newcommand{\kLS}{k_{\mathrm{LS}}}
\newcommand{\kSL}{k_{\mathrm{SL}}}
\newcommand{\kads}{k_{\mathrm{ads}}}
\newcommand{\kadsS}{k_{\mathrm{ads,S}}}
\newcommand{\kdesS}{k_{\mathrm{des,S}}}

\newcommand{\kLL}{k_{\mathrm{LL}}}
\newcommand{\thetaS}{\theta_\mathrm{S}}
\newcommand{\thetaL}{\theta_\mathrm{L}}
\newcommand{\thetaE}{\theta_\mathrm{E}}
\newcommand{\thetaEmax}{\theta_\mathrm{E,max}}
\newcommand{\nLS}{n_\mathrm{LS}}
\newcommand{\nSL}{n_\mathrm{SL}}
\newcommand{\nS}{n_\mathrm{S}}
\newcommand{\nvac}{n_\mathrm{vac}}
\newcommand{\fpr}[1]{\ensuremath{\boldsymbol{#1\!\!:\!\!1}}}
\newcommand{\f}{\ensuremath{f}}
\newcommand{\namegamma}{effective geometric factor}
\newcommand{\namegammas}{effective geometric factors}

\begin{document}

\begin{refsection}

\maketitle

\begin{abstract}
Lying–standing transitions in the first molecular monolayer at organic–inorganic interfaces strongly influence interface dipoles, energy-level alignment, and growth modes, yet their collective kinetics remain difficult to predict. Here, we establish a quantitative adsorbate-to–kinetics relationship for such transitions using first-principles–based kinetic Monte Carlo simulations combined with a mean-field–type coarse-graining strategy. 

Focusing on the prototypical system tetracyanoethylene on Cu(111), we show that the collective transition rate cannot be inferred from any single elementary step but instead emerges from a small set of coupled microscopic processes, including reorientation, adsorption, and diffusion. A local two-step reorientation mechanism captures the diffusion-limited regime, while diffusion of lying molecules accelerates the transition in diffusion-enhanced regimes by sterically suppressing back-reorientation via vacancy–molecule decoupling. This effect is captured by a regime-dependent geometric factor that quantitatively accounts for deviations between single-molecule and collective rate constants.

By systematically varying molecular size and footprint ratio, we demonstrate that geometry provides a powerful intrinsic control parameter. While the collective rate scales approximately proportionally with molecular area, increasing the footprint ratio between lying and standing configurations leads to order-of-magnitude accelerations due to enhanced vacancy creation and diffusion-assisted stabilization. 

Based on these results, we derive an explicit analytical expression for the collective reorientation rate constant that links temperature- and pressure-dependent microscopic rate constants to geometric parameters. The resulting formulation quantitatively reproduces the simulation results across kinetic regimes and provides transferable design principles for engineering lying–standing transition timescales at organic–inorganic interfaces.
\end{abstract}

\section{Introduction}

Organic-inorganic interfaces exhibit a rich variety of structural motifs that strongly influence, for example, their electronic,\cite{ishii_energy_1999,braun_energy-level_2009,hwang_energetics_2009}, thermal,\cite{david_structure_2012} and optical \cite{yokoyama_molecular_2011,baldo_high-efficiency_2000,tiago_ab_2003} properties. Among these, transitions from lying to standing molecular orientations in the first monolayer represent a particularly interesting phenomenon, as these can alter the interface dipole and energy-level alignment. \cite{duhm_orientation-dependent_2008,jeindl_how_2022,chen_energy-level_2019}.

Following Ostwald’s rule of stages, deposition typically initially results in a low-coverage, flat-lying structure.\cite{ostwald_studien_1897} From a thermodynamic perspective, however, often more densely packed upright-standing structures are more stable.\cite{egger_charge_2020,wachter_phase_2025, werkovits_kinetic_2024} Thus, upon deposition of sufficient material, a transformation from the flat-lying into a standing structure is expected to occur. Interestingly, experimentally  such lying-standing transitions have been observed only for a few conjugated molecules (e.g. ref~\cite{broker_density-dependent_2010,hofmann_orientation-dependent_2017,erley_spectroscopic_1987,egger_charge_2020,werkovits_toward_2022,werkovits_kinetic_2024,niederreiter_interplay_2023}), likely because under typical deposition conditions the initial, flat-lying monolayer becomes kinetically trapped.\cite{werkovits_kinetic_2024}
From both a technological and a scientific perspective, being able to engineer the timescale of these phase transitions is of fundamental importance. Predictive control over these processes would enable molecular designs that either deliberately suppress reorientation – allowing to use metastable phases – or promote it sufficiently fast to avoid sequent phase transitions, yielding a stable phase. Both scenarios are crucial for optimizing organic electronic materials.

Unfortunately, presently there are hardly structure-to-property relationships, or in this context more specific adsorbate-to-kinetics relationships, that allow us to predict the timescale at which this phase transition occurs based on adsorbate characteristics, such as energetics and geometry. The main difficulty in determining these timescales occurs from the fact that these transitions involve multiple different processes, possibly including adsorption, desorption, diffusion, and molecular reorientation.  Although for individual molecules, there are often clear design principles how to affect each of these processes separately,\cite{rosei_properties_2003,ruiz_density-functional_2016,arefi_design_2022} in a phase transition they all affect each other. This can lead to collective phase transition rate constants ($\kLScol$) that differ substantially from the rate constants for the individual processes.

The main aim of this work is, thus, twofold: First, we use a representative system (tetracyanoethylene on copper, see below) to establish the functional relationship between the rate constants of the individual-molecule processes and the phase transition. Second, we explore how (and why) this relationship is affected when changing the geometry of the molecule, i.e., the effect of using molecules with different shapes or sizes. Together, these aspects allow us to formulate general principles to control the speed of lying–standing transitions at organic-inorganic interfaces. 

\section{Results and Discussion}

\subsection{Collective Reorientation Kinetics of the Reference System}
\begin{figure}[htb!]
  \centering
  \includegraphics[width=1\textwidth]{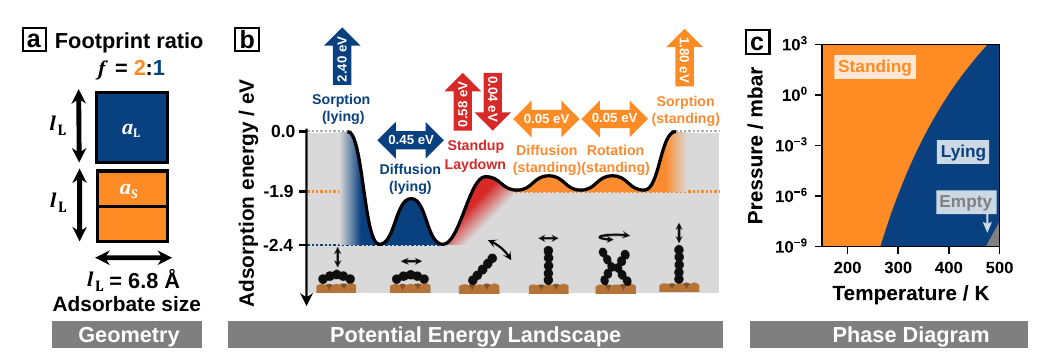}
  \caption{Reference model system illustrating (a) geometry, (b) single-molecule energetics, and (c) thermodynamic stability. (a) Molecular geometry: the planar adsorbate occupies a square footprint of side length $l=\SI{6.8}{\angstrom}$ and area $a_\mathrm{L}=l_L^2$ in the lying orientation (blue), while the standing orientation occupies a smaller area $a_\mathrm{S}$, such that two standing molecules fit into $a_\mathrm{L}$. (b) Schematic potential energy landscape showing adsorption energies and activation barriers for diffusion, reorientation, rotation, and desorption of individual molecules. All on-surface processes are assigned a common attempt frequency of $\SI{1e12}{\per\second}$ (see \nameref{sec:p3_methods}). (c) Thermodynamic phase diagram indicating the orientation with lowest Gibbs free energy of adsorption per area. Orange regions indicate thermodynamically stable standing molecules, blue regions lying molecules, and grey regions conditions under which the empty surface is thermodynamically favored.}
  \label{fig:pub3_fig1}
\end{figure}

The purpose of this section is to establish a quantitative description of the lying-standing phase transition for a specific reference system. Here, we select tetracyanoethylene (TCNE) on Cu(111) as a representative model system for lying-standing transitions, due to the strong experimental and theoretical body of work available in the literature.\cite{erley_spectroscopic_1987,egger_charge_2020,werkovits_toward_2022,johannes_j_cartus_polymorphism_2023,niederreiter_interplay_2023,werkovits_kinetic_2024,wachter_phase_2025}
For this model system, the upright standing phase is thermodynamically stable over a large region of pressures and temperatures,\cite{egger_charge_2020,johannes_j_cartus_polymorphism_2023,werkovits_kinetic_2024,wachter_phase_2025} as shown in Figure~\ref{fig:pub3_fig1}c (orange region).

To determine the phase transition rate constant $\kLScol$, we perform kinetic Monte Carlo simulations across a wide range of temperature-pressure conditions where lying-standing transitions occur. The adsorbate is modelled as a two-dimensional object on a square grid. As visualized in Figure~\ref{fig:pub3_fig1}a, the molecule can be either in a lying configuration (blue) or in an upright standing configuration (orange). For our reference system, the lying configuration occupies a region of $\SI{6.8}{\angstrom}\times\SI{6.8}{\angstrom}$, while the standing configuration occupies half that space, $\SI{6.8}{\angstrom}\times\SI{3.4}{\angstrom}$. Within our simulation, we consider reorientation between the two configurations, as well as adsorption, desorption, and diffusion of each configuration, using the relative energies and barriers shown in Figure~\ref{fig:pub3_fig1}b. Importantly, reorientation from standing to lying can only take place if there is free space available, i.e. if the adjacent site is not occupied by another molecule. Further details are given in the \nameref{sec:p3_methods} section.

\FloatBarrier

To probe the kinetics of the collective lying-standing phase transition, we start with a system that initially consists completely of flat-lying molecules. We then track the time evolution of lying and standing molecules under conditions where the standing phase is thermodynamically favored. This setup reflects experimentally relevant growth scenarios, in which deposition first produces a metastable lying phase that subsequently undergoes a collective reorientation.

To extract a compact measure of the collective reorientation kinetics, we employ the \textit{irreversible power-law two-state approximation} (\textit{IPL2SA}, see \nameref{sec:p3_methods}). In this approximation, the detailed spatial evolution of the adsorbate layer is not resolved explicitly. Rather, all effective influences are captured implicitly through two effective parameters obtained by fitting the simulated coverage-time profiles. Equation~\ref{eq:pub3_1} states how the coverage fraction of standing molecules $\thetaS$ changes based on the coverage fraction of lying molecules $\thetaL$ and defines the collective rate constant $\kLScol$ and the apparent reaction order $\alpha$, simply referred to as “reaction order” hereafter.
\begin{equation}
  \frac{\mathrm{d}\thetaS}{\mathrm{d}t} = \kLScol \cdot \thetaL^{\alpha}
  \label{eq:pub3_1}
\end{equation}

The reaction order quantifies how strongly the transition rate depends on the current coverage of lying molecules. Importantly, $\alpha$ does not correspond to molecularity in the classical sense (e.g., unimolecular or bimolecular reactions). Instead, it reflects emergent collective behavior, measuring how the evolving surface structure modulates the probability that a single-molecule reorientation becomes stabilized. Values of $\alpha$~<~1 indicate that the reaction is autocatalytic (i.e., the formation of standing molecules accelerates the reaction), while $\alpha$~>~1 indicates a self-inhibitory effect of the reaction product on the reaction rate. 

\begin{figure}[htb!]
  \centering
  \includegraphics[width=1\textwidth]{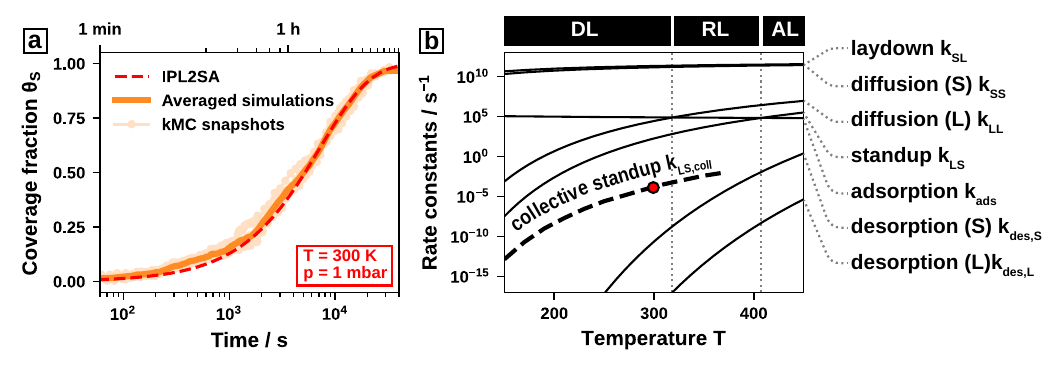}
  \caption{(a) Coverage fraction of standing adsorbates $\thetaS$ as a function of time at $T=\SI{300}{\kelvin}$ and $p~=~\SI{1}{\milli\bar}$. Snapshots of five independent kMC simulations (faded markers) are averaged to obtain a statistically robust coverage profile (orange solid line) for fitting via the IPL2SA (dashed red line). (b) Temperature-dependence of the collective rate constant for the lying-standing transition (dashed) relative to the single-molecule rate constants (solid, with annotated microscopic processes) for a pressure of $p=\SI{1}{\milli\bar}$. The intersections of single-molecule rate constants indicate changes of kinetic regimes between diffusion-limited (DL), reorientation-limited (RL) and adsorption-limited (AL). The red marker highlights the collective rate constant obtained from the IPL2SA fit at $T=\SI{300}{\kelvin}$ and $p~=~\SI{1}{\milli\bar}$ (panel a).}
  \label{fig:pub3_fig2}
\end{figure}

To illustrate typical shapes of the simulated coverage-time profiles and to demonstrate the quality of the IPL2SA approximation, we show the time evolution of the reference system at near-ambient temperature and moderate pressure ($T=\SI{300}{\kelvin}$, $p=\SI{1}{\milli\bar}$) in Figure~\ref{fig:pub3_fig2}a. To obtain statistically robust lying-standing transitions, five independent simulation runs are performed. The faded orange markers represent snapshots from these runs, while their average yields the smooth coverage-time profile shown as the solid orange line. The fit yields $\alpha$~=~1.1, which indicates a slight self-inhibition due to steric constraints, and $\kLScol$~=~\SI{1.4e-4}{\per\second}. At first glance, one might expect the collective rate constant to be governed by the slowest elementary process directly involved in the transition—under these conditions, the single-molecule reorientation rate $\kLS$. However, direct comparison of the rate constants (Figure~\ref{fig:pub3_fig2}b) reveals that $\kLScol$ is, in fact, about six orders of magnitude smaller than $\kLS$.

\begin{figure}[htb!]
  \centering
  \includegraphics[width=1\textwidth]{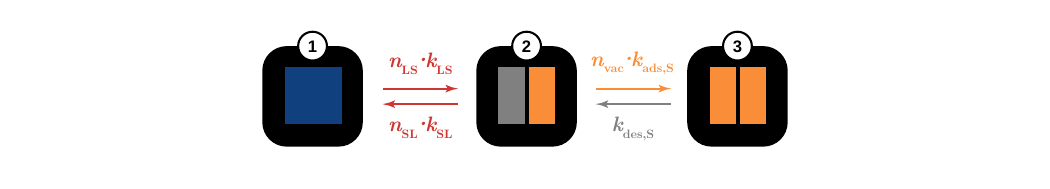}
  \caption{Dominant reaction channel illustrating the local two-step reorientation mechanism in the absence of diffusion. Following reorientation (State~1 $\rightarrow$ State~2), the system either returns to the lying configuration via back-reorientation or proceeds by adsorption to complete the effective reorientation (State~3). The local reorientation event is treated in isolation from the surrounding adsorbates (black area), whose orientations are assumed arbitrary and whose coupling to the transition site is neglected. Annotated arrows indicate directions and total rate constants ($n_{\mathrm{i} \rightarrow \mathrm{j}} \cdot k_{\mathrm{i} \rightarrow \mathrm{j}}$) of single-molecule transitions between microstates $i$ and $j$. Blue: lying; orange: standing; grey: empty.}
  \label{fig:pub3_fig3}
\end{figure}

To understand the deviation between collective and single-molecule rate constants, we first introduce a simplified local-transition model of the phase transformation that neglects diffusion, which is incorporated later in this work. This model treats the collective lying–standing transition as an effectively decoupled sequence of single-molecule processes. 
Specifically, the local reorientation event is considered in isolation from the surrounding adsorbates. 
In this sense, the model represents a reduced reaction network that focuses solely on the transition pathway of a single lying molecule embedded in an otherwise unspecified environment (illustrated schematically in Figure~\ref{fig:pub3_fig3}).

Within this reduced local-transition network, the phase transformation proceeds via a two-step mechanism.\cite{werkovits_kinetic_2024} 
First, an individual molecule reorients from a lying orientation (State~1 in Figure~\ref{fig:pub3_fig3}) to a standing orientation while creating an adjacent vacancy (State~2), occurring with rate constant $\kLS$. 
The reverse process, i.e.\ falling over, proceeds with rate constant $\kSL$. 
Once State~2 is formed, the vacancy is filled by adsorption of a standing molecule (State~3) with rate constant $\kadsS$. 
This second step can only be reversed by desorption of a standing molecule with rate constant $\kdesS$. 
For large organic molecules, desorption is typically very slow (see below), rendering the two-step sequence locally quasi-irreversible.

Thus, at the global level, the collective phase transition can be mapped onto this reduced network representation of a single-molecule transition pathway, as illustrated in Figure~\ref{fig:pub3_fig3}.

The evolution equation follows directly from the propensity formulation of stochastic chemical kinetics.\cite{gillespie_general_1976} The propensity describes how likely a specific reaction is to occur in the next infinitesimal time interval.
For each elementary reaction channel $\mu$, the propensity is written as
\begin{equation}
a_\mu(\mathbf{x}) 
= k_\mu \, h_\mu(\mathbf{x})
= k_\mu \, p_i \, n_i ,
\end{equation}
where $k_\mu$ is the microscopic rate constant. 
Here, $h_\mu(\mathbf{x})$ corresponds to the number of available configurations susceptible to $\mu$, 
which in the present case factorizes into the probability $p_i$ of finding the system in the relevant local configuration 
and the number $n_i$ of possible transitions accessible from this configuration, 
consistent with a mean-field mass–action description. 

The rate equation is then obtained as the net balance of gain and loss propensities of State~3,
\begin{equation}
\frac{\mathrm{d} p_3}{\mathrm{d} t}
 = p_2(t) \cdot \nvac \cdot \kadsS 
 - p_3(t) \cdot \nS \cdot \kdesS.
 \label{eq:pub3_rate_equ_local_orig}
\end{equation}
Here, $p_3$ denotes the probability that a local region is in the fully converted standing state (State~3), 
while $p_2$ is the probability of the intermediate configuration consisting of a standing molecule 
with an adjacent vacancy (State~2). 
The quantity $\nvac$ represents the number of adsorption-enabled vacancies created upon reorientation, 
i.e.\ vacancies large enough to accommodate a standing molecule. 
Conversely, $\nS$ denotes the number of standing molecules within a reference cell of the size of a lying molecule, 
and thus the number of possible desorption events from State~3.

Because every configuration produced by a reorientation event is rapidly consumed either by back-reorientation to the lying state (State~1) or by adsorption (leading to State~3), the intermediate State~2 is short-lived. Consequently, $p_2$ does not accumulate and remains small compared to $p_1$ and $p_3$. Under these conditions, the steady-state approximation \cite{gold_iupac_2019} applies, which assumes that the intermediate relaxes on a much faster timescale than the overall phase transformation. The effective collective transition rate can therefore be expressed solely in terms of single-molecule rate constants and transition multiplicities.

\begin{equation}
\frac{\mathrm{d} p_2}{\mathrm{d} t} \approx 0
\quad \Rightarrow \quad
p_2 = \frac{\nLS \cdot \kLS \cdot p_1 + \nS \cdot \kdesS \cdot p_3}{\nSL \cdot \kSL + \nvac \cdot \kadsS}
\label{eq:pub3_3}
\end{equation}

The quantity $\nLS$ denotes the number of symmetry-equivalent possibilities in State~1 ($p_1$) that allow a molecule to stand up, while $\nSL$ gives the number of possibilities for falling back to the lying configuration (State~3 with $p_3$). Since by construction $\kdesS \ll \kLS$ (desorption is energetically more costly than reorientation; see Figure~\ref{fig:pub3_fig1}a), and since $\kSL$ is much faster than adsorption $\kadsS$ (Figure~\ref{fig:pub3_fig2}b), combining Equations~\ref{eq:pub3_3} and \ref{eq:pub3_rate_equ_local_orig} yields

\begin{equation}
\frac{\mathrm{d} p_3}{\mathrm{d} t}
= \gamma \cdot \frac{\kLS}{\kSL}\cdot \kadsS \cdot p_1,
\quad \text{with }
\gamma = \frac{\nvac \cdot \nLS}{\nSL}.
\label{eq:pub3_rate_equ_local}
\end{equation}

The transition multiplicities are thus condensed into the \namegamma~$\gamma$. In the present model, a molecule can reorient in four directions ($\nLS = 4$) by creating one vacancy ($\nvac = 1$), whereas only one pathway exists for reverting to the lying state ($\nSL = 1$), resulting in $\gamma = 4$.

All geometric contributions of the local two-step reorientation are therefore captured by the prefactor $\gamma$. Using the mean-field equivalence between configuration probabilities and coverage fractions ($p_1 = \thetaL$, $p_2 \approx 0$ $p_3 = \thetaS$), Equation~\ref{eq:pub3_rate_equ_local} can be recasted in the IPL2SA form (Equation~\ref{eq:pub3_1}),

\begin{equation}
\frac{\mathrm{d} \thetaS}{\mathrm{d} t}
= \kLScol \cdot \thetaL^{\alpha},
\quad \text{with }
\kLScol = \gamma \cdot \frac{\kLS}{\kSL}\cdot \kadsS.
\label{eq:pub3_4}
\end{equation}

The apparent reaction order $\alpha$ is allowed to deviate from unity to generalize the model beyond a strictly first-order description and may effectively account for coverage-dependent or additional geometric influences. Equation~\ref{eq:pub3_4} thus provides a theory-based decomposition of the collective rate constant $\kLScol$, with the dominant geometric contribution entering explicitly through $\gamma$.

\begin{figure}[htb!]
  \centering
  \includegraphics[width=1\textwidth]{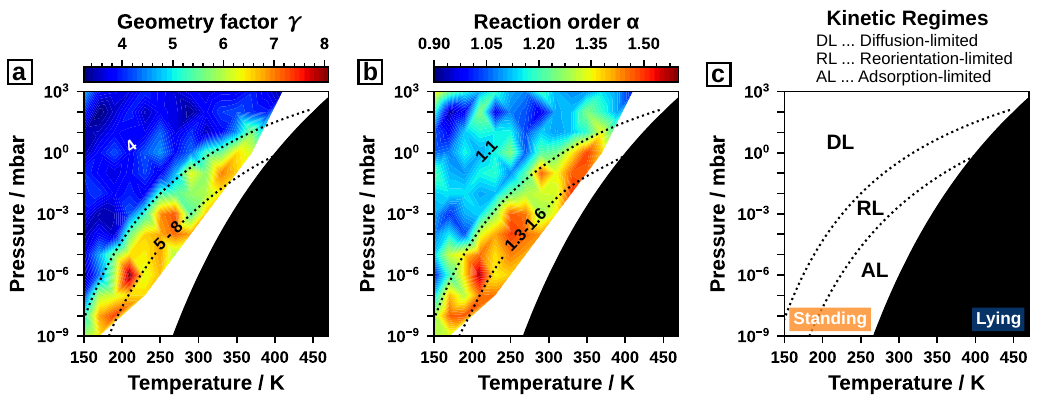}
  \caption{Pressure–temperature diagrams of (a) the \namegamma~$\gamma$, (b) the apparent reaction order $\alpha$, and (c) the kinetic regimes governing the collective reorientation. (a) $\gamma$ quantifies by which factor geometric and steric effects modify the effective transition rate beyond the basis approximation $\kLS\cdot\kSL^{-1}\cdot\kadsS$ from Equation~\ref{eq:pub3_4}. (b) $\alpha>1$ indicate self-inhibitory effects. (c) Temperature-pressure regions of diffusion-limited (DL), reorientation-limited (RL), and adsorption-limited (AL) kinetic regimes; their boundaries are indicated by dotted lines also in panels (a) and (b). The thermodynamic phase diagram is shown in the background for reference, with regions where standing and lying molecules are thermodynamically stable displayed in white and black.}
  \label{fig:pub3_fig4}
\end{figure}

To validate this formulation, we extract the \namegamma~$\gamma$ from IPL2SA fits to the kMC simulations and analyze its dependence on temperature and pressure (Figure~\ref{fig:pub3_fig4}a). Far from the phase boundary between lying and standing molecules (cf.\ Figure~\ref{fig:pub3_fig1}c), i.e.\ at high pressures and low temperatures, $\gamma$ is close to 4, consistent with the symmetry argument. Closer to the phase boundary, however, $\gamma$ increases noticeably. This trend appears counterintuitive, as the increasing thermodynamic stability of the standing phase away from the boundary would, according to the Bell–Evans–Polanyi principle, \cite{evans_inertia_1938,bligaard_bronstedevanspolanyi_2004} suggest the opposite behavior.

The observed increase in $\gamma$ indicates that at higher temperatures, additional processes beyond the local effective two-step mechanism (reorientation and adsorption) contribute to the collective transition. This interpretation is supported by the behavior of the reaction order $\alpha$ as a function of pressure and temperature (Figure~\ref{fig:pub3_fig4}b). At low temperatures, $\alpha$ is close to unity, whereas in regions where $\gamma > 4$, it increases to values of up to $\approx 2$. For both $\alpha$ and $\gamma$, the transition from low to high values occurs rather abruptly. Consistently, the collective rate constants $\kLScol$ cannot be described over the entire temperature–pressure range by a single effective Arrhenius expression with constant attempt frequency and barrier. Instead, as visualized in Section~\ref{sec:p3_SI_arrh} of the \nameref{sec:p3_SI}, both parameters exhibit a temperature and pressure dependence, which becomes even more pronounced for increasing footprint ratios (as investigated later).

This behavior naturally motivates the introduction of kinetic regimes. Within a given regime, the same hierarchy of single-molecule processes governs the collective reorientation dynamics. To illustrate this concept, Figure~\ref{fig:pub3_fig2}b shows the temperature dependence of the single-molecule rate constants at a representative pressure of $p=\SI{1}{\milli\bar}$. Notably, among the considered single-molecule processes, only the adsorption rate constant depends on pressure. At fixed pressure, intersections between rate-constant curves divide the temperature axis into distinct intervals in which the ordering of the underlying molecular processes remains unchanged. For the reference system, this yields three major kinetic regimes: In the diffusion-limited (\textbf{DL}) regime, the rate constants for diffusion of lying molecules and reorientation from lying to standing are both slower than adsorption. In the reorientation-limited regime (\textbf{RL}), diffusion is faster than the adsorption process and reorientation slower than both. Finally, in the adsorption-limited regime (\textbf{AL}), also the reorientation process is faster than the adsorption process. The corresponding regime-dependent temperature intervals can be mapped onto a diagram, which yields contiguous regions in the  diagram shown in Figure~\ref{fig:pub3_fig4}c. Comparing the variations of $\gamma$ and $\alpha$ with the boundaries of these kinetic regimes (shown as dotted lines in Figure~\ref{fig:pub3_fig4}a and b) shows good agreement, corroborating the interpretation that additional processes play a role in accelerating the phase transition.

A defining characteristic of our system is that molecules which are standing upright experience a much lower diffusion barrier than flat-lying molecules (compare Figure~\ref{fig:pub3_fig1}a). Consequently, even at low temperatures (i.e. in the \textbf{DL} regime), standing molecules are very mobile, while flat-lying molecules essentially remain frozen in place. However, in our reference system, the diffusion of standing molecules to vacancies (which is only possible in State 2) leads to a symmetry-equivalent state. Hence, here it has virtually no impact on the \namegamma~$\gamma$ at all.

\begin{figure}[htb!]
  \centering
  \includegraphics[width=1\textwidth]{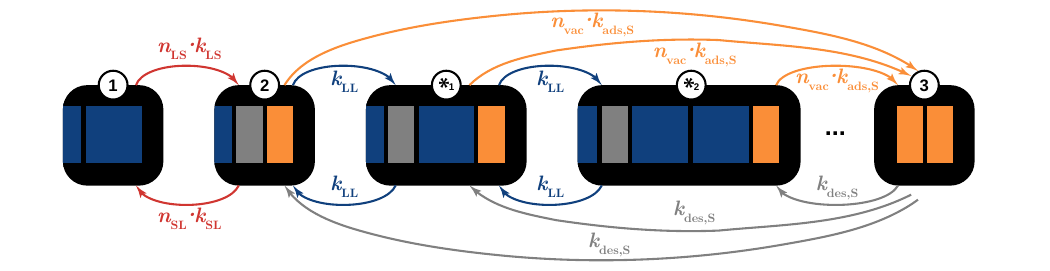}
  \caption{Reaction channel illustrating vacancy–molecule decoupling in the presence of lying-molecule diffusion. After reorientation (State~1 $\rightarrow$ State~2), diffusion stabilizes the standing molecule (State~$\ast_1$ or States~$\ast_i$ for consecutive possible diffusions), enabling completion of the effective reorientation by adsorption (State~3, here displayed simplified and generalized consistent with Figure~\ref{fig:pub3_fig3}). Annotated arrows indicate directions and total rate constants ($n_{\mathrm{i} \rightarrow \mathrm{j}} \cdot k_{\mathrm{i} \rightarrow \mathrm{j}}$) of single-molecule transitions between microstates $i$ and $j$, whereas with Equation~\ref{eq:pub3_pvacS_main} the effective total rate constant of State~2 $\rightarrow$ 1 becomes $\kSL \cdot \nSL \cdot p_{\mathrm{vac+S}}$ after incorporating effects of diffusion stabilization. Blue: lying, orange: standing, grey: empty, black: arbitrary surrounding.}
  \label{fig:pub3_fig5}
\end{figure}

Conversely, in \textbf{RL} and \textbf{AL} regimes, also flat-lying molecules become mobile, i.e. the diffusion becomes faster than adsorption of additional molecules. In this regime, once a standing molecule with an adjacent vacancy is formed (State 2), lying molecules from the neighboring cell can move into the empty space, essentially dislocating the vacancy by two lattice sites (State $\ast_1$). If further diffusion steps are possible, the vacancy itself can migrate even farther (States~$\ast_i$). While the vacancy still allows an additional molecule to adsorb, the upright-standing molecule is now sterically forbidden to reorient back into the lying configuration (State 1) again. This extended process is indicated in Figure~\ref{fig:pub3_fig5}.

We now incorporate the effective stabilization of previously reoriented standing molecules induced by diffusion of lying molecules (States~$\ast_i \rightarrow$ State~3) into the theory-derived \namegamma~$\gamma$, which in Equation~\ref{eq:pub3_rate_equ_local} arises solely from the local two-step reorientation mechanism. To this end, the diffusion reaction channels shown in Figure~\ref{fig:pub3_fig5} are mapped onto the local two-step reorientation mechanism of Figure~\ref{fig:pub3_fig3} by explicitly accounting for the probability that back-reorientation remains possible.

Back-reorientation (S$\rightarrow$L) requires that a vacancy exists and that this vacancy is adjacent to the standing molecule. The corresponding probability factorizes into the probability of vacancy formation and the conditional probability that this vacancy is found adjacent to the standing molecule. This probability is reduced if diffusion of lying molecules removes  the vacancy from the standing molecule before back-reorientation occurs.

As derived in Section~\ref{sec:p3_SI_vacstab} of the Supporting Information, the probability that a vacancy remains adjacent to a standing molecule is

\begin{equation}
  p_{\mathrm{vac+S}}
  =
  1 - \frac{\kLL}{\kLL + \nvac \kadsS + \omega \kLL},
  \label{eq:pub3_pvacS_main}
\end{equation}

which expresses the competition between vacancy blocking via lying diffusion ($\kLL$) and vacancy consumption via adsorption ($\nvac \kadsS$) or further diffusion events (captured by $\omega \kLL$).

The factor $\omega$ is an effective multiplicity parameter that quantifies how many additional diffusion pathways exist which generate configurations in which the vacancy is no longer adjacent to the originally reoriented standing molecule. Physically, $\omega$ accounts for the fact that once a lying molecule diffuses into the vacancy created during reorientation, subsequent diffusion steps can further displace the vacancy and thereby reduce the probability that the standing molecule can fall back. In this sense, $\omega$ measures how (ir)reversibly diffusion removes the vacancy from the standing molecule.

For the reference system a single diffusion step of a neighboring lying molecule necessarily creates a new vacancy elsewhere. However, this new vacancy is no longer adjacent to the original standing molecule. Thus, although the number of vacancies is conserved, the spatial correlation between vacancy and standing molecule is lost. $\omega = 1$ corresponds to the case where diffusion creates, on average, one additional equivalent configuration that competes with back-reorientation. For fast diffusion ($\kLL \gg \kadsS$), the expression in Equation~\ref{eq:pub3_pvacS_main} then approaches

\[
p_{\mathrm{vac+S}} \;\longrightarrow\; 1-\frac{1}{1+\omega}.
\]

For $\omega=1$, this yields $p_{\mathrm{vac+S}} = \tfrac{1}{2}$. Physically, this means that under rapid lying diffusion the vacancy is equally likely to remain adjacent to the standing molecule or to become displaced to a neighboring configuration. Back-reorientation therefore remains possible only in roughly half of the microscopic realizations, effectively suppressing the reverse process by a factor of two.

Incorporating this probability into the effective back-reorientation propensity yields the extended geometric prefactor

\begin{equation}
  \gamma
  =
  \frac{\nvac \cdot \nLS}
       {\nSL \cdot 
        \underbrace{\left(
        1 - \frac{\kLL}{\kLL + \nvac \kadsS + \omega \kLL}
        \right)}_{p_{\mathrm{vac+S}}}
       },
  \label{eq:pub3_gamma_ext_main}
\end{equation}

which generalizes the purely local two-step approximation of Equation~\ref{eq:pub3_rate_equ_local}.

In the diffusion-limited regime ($\kLL < \kadsS$), one obtains $p_{\mathrm{vac+S}} \approx 1$, and diffusion does not affect the kinetics. In contrast, for fast lying diffusion ($\kLL > \kadsS$) and $\omega=1$, the probability converges toward $p_{\mathrm{vac+S}} = \tfrac{1}{2}$, which yields $\gamma = 8$ for the present reference system. This behavior is qualitatively consistent with Figure~\ref{fig:pub3_fig4}a, where $\gamma$ assumes values from \numrange{5}{8} in regimes where diffusion is not limiting (\textbf{RL} and \textbf{AL} regimes). Diffusion of lying molecules therefore accelerates the collective phase transition not by enhancing the forward reorientation step, but by sterically suppressing the reverse process and thereby stabilizing newly formed standing molecules.

Furthermore, whether this diffusion process becomes relevant depends on the likelihood that the neighboring unit cell is indeed occupied by a lying molecule (State~1) rather than two standing molecules (State~3). Thus, the observed acceleration is self-inhibitory, i.e. the formation of the product inhibits the acceleration of the formation of further product, which is consistent with the observation that $\alpha > 1$ in this region (see Figure\ref{fig:pub3_fig4}b).

While the theoretical analysis predicts $\gamma = 4$ for the diffusion-limited (\textbf{DL}) regime and $\gamma = 8$ for the remaining regimes, the kMC simulations systematically yield slightly smaller values across the full parameter range, with $\gamma \approx \numrange{3}{4}$ in the \textbf{DL} regime and $\gamma \approx \numrange{5}{8}$ otherwise. This systematic deviation can be traced back to an additional escape channel from State~3 in the reduced reaction network, shown in Figure~\ref{fig:pub3_fig3}, that is not included so far. Specifically, when a unit cell in State~1 is adjacent to a unit cell in State~3, the vacancy created by the reorientation of the lying molecule can facilitate a renewed falling-over event of the standing molecule. In detail, this mechanism, referred to as \textit{neighbor-induced back-reorientation}, effectively increases the apparent single-molecule reorientation rate $\kLS$ by up to a factor of two, weighted by the probability of finding standing molecules adjacent to lying ones. Therefore, the transition becomes progressively inhibitive for higher coverage fractions of standing molecules, and slightly decelerates the overall phase transition. Importantly, it closes the gap between the kMC-retrieved and theoretical \namegamma~$\gamma$, and therefore as well between the respective formulations of the collective rate constant. 



\subsection{Influence of adsorbate geometry}

Having established how the collective lying–standing transition of the reference system can be rationalized in terms of microscopic rate constants and an effective geometric factor $\gamma$, we now address the second central question of this work: How molecular geometry modifies collective kinetics. In the preceding section, $\gamma$ was shown to encode steric multiplicities and diffusion-induced stabilization effects via the probability $p_{\mathrm{vac+S}}$. Consequently, geometric variations are expected to directly affect the balance between forward reorientation, vacancy stabilization, and back-reorientation.

Specifically, we investigate how changes in molecular footprint and packing constraints influence the effective geometric factor $\gamma$. While external growth parameters such as temperature and pressure primarily shift the system between kinetic regimes, molecular geometry modifies the microscopic transition multiplicities ($\nLS$, $\nSL$, $\nvac$) and the efficiency of vacancy–molecule decoupling. Geometry therefore acts as an intrinsic control parameter that can amplify or suppress diffusion-induced stabilization mechanisms. By systematically varying footprint ratios and molecular sizes, we aim to identify how steric constraints translate into quantitative modifications of the collective rate law and thereby establish transferable structure–kinetics relations for lying–standing transitions at organic–inorganic interfaces.

\begin{figure}[htb!]
  \centering
  \includegraphics[width=0.5\textwidth]{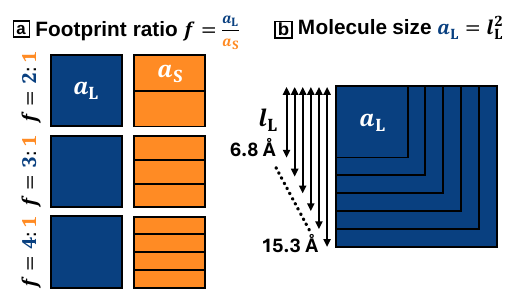}
  \caption{Schematics of geometric variation: (a) footprint ratio and (b) molecule size. (a) The footprint ratio  is defined as ratio of the footprint area of lying and standing adsorbates,  and  respectively. This means that  molecules in the standing orientation fit in the footprint area of a molecule in lying orientation.  is varied from 2 (reference) to 3 and 4. (b) The size of the planar, quadratic molecule is defined by its area , where  denotes the side length of the molecule. For each footprint ratio  is varied from $\SI{6.8}{\angstrom}$ (reference) to $\SI{15.3}{\angstrom}$.}
  \label{fig:pub3_fig6}
\end{figure}

To isolate geometric from energetic effects, the adsorbate geometry is systematically varied while the underlying energetic landscape remains fixed to that of the reference system, as shown in Figure~\ref{fig:pub3_fig1}b. Two geometric characteristics are explored – the overall molecular size and the footprint ratio between lying and standing adsorption positions, as illustrated in Figure~\ref{fig:pub3_fig6}.  Details of the lattice representation, systematics of size variation and geometric implementation are provided in Section \ref{sec:p3_SI1} of the Supporting Information. We note that although the underlying single-molecule energetic landscape is kept fixed, increasing the footprint ratio modifies the thermodynamic phase diagram: Because the adsorption energy per standing molecule is unchanged while standing molecules can pack more densely, the adsorption energy per unit area becomes more favorable, shifting the thermodynamic lying-standing phase boundary toward higher temperatures, as shown in SI Section~\ref{sec:p3_SI_kMCsampling}.

\begin{figure}[htb!]
  \centering
  \includegraphics[width=0.5\textwidth]{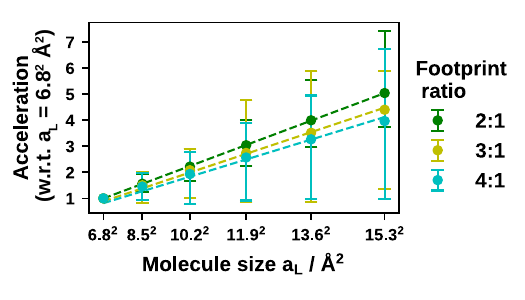}
  \caption{Adsorbate-size–induced acceleration factor of the collective rate constant, defined as $\kLScol(a_\mathrm{L}) / \kLScol(a_{\mathrm{L,ref}})$, shown as a function of molecular area $a_\mathrm{L} = l_{\mathrm{L}}^{2}$. The reference size corresponds to $l_{\mathrm{L}} = \SI{6.8}{\angstrom}$. Results are displayed separately for footprint ratios $f = 2{:}1$ (green), $3{:}1$ (yellow), and $4{:}1$ (cyan). Symbols indicate mean acceleration factors averaged over all $(T,p)$ points; error bars denote the minimum–maximum range. The dashed lines shows a fitted proportionality for increasing footprint ratios, respectively.
}
  \label{fig:pub3_fig7}
\end{figure}

In our simulations, the molecular size is systematically varied by increasing the lying footprint area $a_{\mathrm{L}} = l_{\mathrm{L}}^{2}$ from the reference value of $(\SI{6.8}{\angstrom})^{2}$ up to $(\SI{15.3}{\angstrom})^{2}$. Keeping the adsorption rate constant $k_{\mathrm{ads}}$ and the temperature fixed, we find that larger molecules undergo faster lying-standing phase transitions. As shown in Figure~\ref{fig:pub3_fig7}, this dependence is almost perfectly proportional ($\kLScol \propto a_\mathrm{L}$) propagating the size-dependency from the adsorption rate constant to the collective rate constant. This behavior is readily understood: For larger molecules, each single-molecule process affects a larger fraction of the total surface area, such that fewer  events are required to transform the entire layer. Notably, this scaling scarcely depends on the footprint ratio and is independent of other molecular properties by construction. We note in passing that larger molecules generally also have a proportionally larger mass; consequently, maintaining the same adsorption rate for different molecules requires a correspondingly higher pressure (see Equation~\ref{eq:pub3_kads} in the \nameref{sec:p3_methods}).

\begin{figure}[htb!]
  \centering
  \includegraphics[width=1\textwidth]{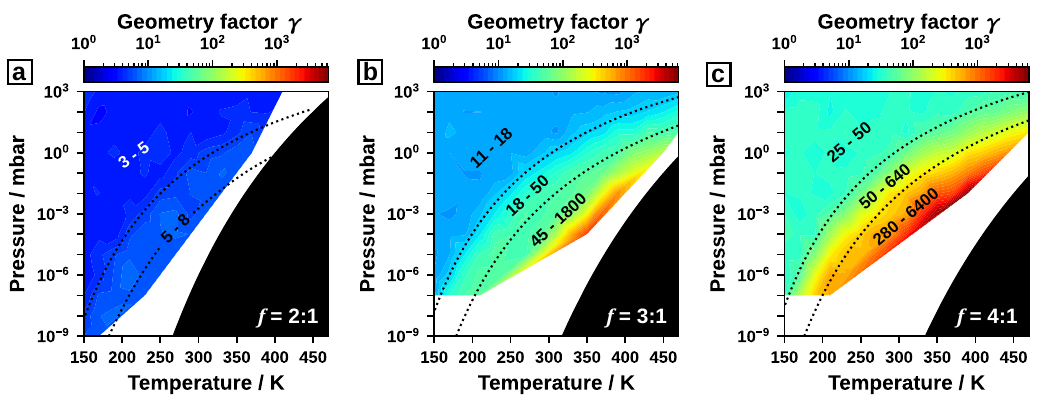}
  \caption{
  Pressure–temperature diagrams of the \namegamma~$\gamma$ for model systems with increasing footprint ratios: (a) $\fpr{2}$, (b) $\fpr{3}$, and (c) $\fpr{4}$. $\gamma$ quantifies the steric enhancement of the effective transition rate beyond the local two-step expression $\kLS\cdot\kSL^{-1}\cdot\kadsS$ (Equation~\ref{eq:pub3_4}). An identical color scale is used for all panels to enable direct comparison. Boundaries between diffusion-limited (\textbf{DL}), reorientation-limited (\textbf{RL}), and adsorption-limited (\textbf{AL}) regimes are indicated by dotted lines, and the corresponding $\gamma$ ranges are annotated. The thermodynamic phase diagram is shown in the background for reference, highlighting regions where standing (white) or lying (black) molecules are thermodynamically stable. 
  \label{fig:pub3_fig8}
  }
\end{figure}

As the other prominent geometric handle, we focus on the footprint ratio. A model system $\fpr{f}$ with a footprint ratio $\f$ denotes that $f$ standing molecules can occupy the same surface area as one lying molecule. Such ratios naturally arise for conjugated backbones, from relatively compact systems such as benzene-like molecules ($f \approx 2$) to increasingly extended backbones such as naphtalene- ($f \approx 3$) or anthracene-like ($f \approx 4$) molecules with larger $\f$.\footnote{Here, we consider the case in which standing molecules would contact the substrate via their short molecular edge.} As depicted in Figure~\ref{fig:pub3_fig6}a, in our simulation we use model systems with footprint ratios of $\f=\numlist{2;3;4}$. Figure~\ref{fig:pub3_fig8} plots their corresponding \namegammas~$\gamma$, that quantify how strongly spatial effects arising from the adsorbate geometry accelerate collective lying-standing transitions relative to the simplified diffusion-inhibited approximation $\kLS\cdot\kSL^{-1}\cdot\kadsS$ (cf. Equation~\ref{eq:pub3_4}).
\FloatBarrier
\FloatBarrier
Qualitatively, all three systems exhibit a common acceleration behavior: $\gamma$ assumes its smallest value in the diffusion-limited (\textbf{DL}) regime, i.e., at low temperatures and increases substantially upon approaching the corresponding lying–standing phase boundary. This trend highlights consistently a qualitative change in the dominant microscopic mechanisms when transitioning from diffusion-limited to diffusion-enhanced (\textbf{DE}) regime, the latter encompassing both the reorientation-limited (\textbf{RL}) and adsorption-limited (\textbf{AL}) regimes. Quantitatively, however, the magnitude of $\gamma$ depends strongly on the footprint ratio. For the $\fpr{3}$ system, $\gamma$ already reaches values between 10 and 20 in the DL regime, while for $\fpr{4}$ it increases further to approximately 25–50. In the diffusion-enhanced regimes, lying–standing transitions for the $\fpr{3}$ and $\fpr{4}$ systems are accelerated by approximately one to two orders of magnitude compared to their diffusion-limited counterparts. Larger molecular footprints therefore lead to a pronounced, geometry-assisted acceleration of the collective phase-transition rate constant.

This footprint-ratio-dependent enhancement arises from two geometric effects that are already contained in the local two-step reorientation picture. 

First, reorientation of lying molecules with larger footprints generates larger contiguous vacant regions. In terms of the basic reaction channel shown in Figure~\ref{fig:pub3_fig3}, a molecule with footprint ratio $f$ creates $\nvac = f - 1$ vacancies upon reorientation. This directly increases the number of adsorption-enabled sites available for stabilization.

Second, the reoriented standing molecule can diffuse within this enlarged vacancy region. Falling-over events remain sterically restricted to edge positions. However, standing-molecule diffusion allows the molecule to also move to interior adsorption sites, where back-reorientation is sterically suppressed. Since standing diffusion is fast on the relevant time scales, the standing molecule can be assumed to sample the vacancy region quasi-uniformly. As a result, the effective multiplicity for falling-over events is reduced by a factor $\nSL = 2/f$.

Taken together, these two geometric effects yield the collective rate constant

\begin{equation}
  \kLScol = 
  \frac{4 (f-1) \cdot \kLS \cdot \kadsS}
       {\kSL \cdot \dfrac{2}{f} \cdot 
       \left( 1-\dfrac{\kLL}{(1+\omega)\cdot\kLL+ \nvac \cdot \kadsS}\right)}
  =
  2 (f^2-f) 
  \frac{\kLS \cdot \kadsS}
       {\kSL \cdot 
       \left( 1-\dfrac{\kLL}{(1+\omega)\cdot\kLL+ \nvac \cdot \kadsS}\right)},
  \label{eq:pub3_6}
\end{equation}

which generalizes the local two-step expression by explicitly incorporating footprint ratio $f$ and resembles effective geometric factors $\gamma$ in the diffusion-limted regimes (\textbf{DL}).

\FloatBarrier

Beyond the two purely geometric multiplicity effects discussed above, larger footprint ratios additionally enhance vacancy–molecule decoupling mediated by diffusion of lying molecules. This third mechanism is captured by the effective multiplicity factor $\omega$, which quantifies how efficiently diffusion redistributes vacancies away from the originally formed standing molecule.

Physically, $\omega$ measures how many diffusion-mediated configurations effectively compete with adsorption in determining whether the vacancy remains adjacent to the standing molecule. For $\omega = 1$, diffusion generates essentially one additional equivalent configuration.  In this situation, vacancy–standing decoupling remains partially reversible: Although diffusion can temporarily displace the vacancy, the probability of reforming the original adjacency remains significant. For larger footprint ratios and in diffusion-enhanced regimes, however, lying molecules can access multiple intermediate positions within the extended vacancy region. This enables progressive delocalization of the vacancy across the adlayer. 

When diffusion rapidly propagates and redistributes vacancies such that the probability of re-forming the original vacancy–standing adjacency becomes negligible, $\omega$ becomes small. In this limit, vacancy–standing decoupling is effectively irreversible on the timescale of adsorption and back-reorientation, leading to a strong reduction of $p_{\mathrm{vac+S}}$ and consequently to large values of the geometric prefactor $\gamma$.

\begin{figure}[htb!]
  \centering
  \includegraphics[width=1\textwidth]{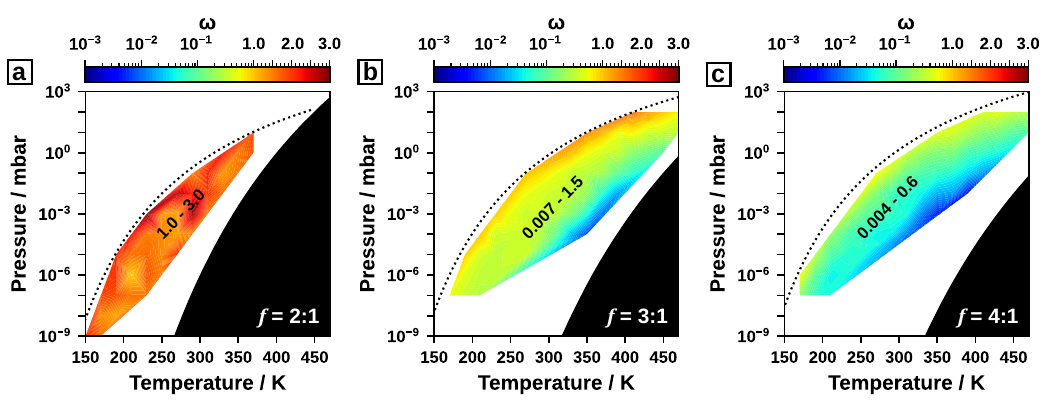}
  \caption{Pressure–temperature diagrams of $\omega$ for model systems with increasing footprint ratios: (a) $\fpr{2}$, (b) $\fpr{3}$, and (c) $\fpr{4}$. $\omega$ quantifies how likely a vacancy–standing decoupling caused by diffusion in the lying orientation remains partially reversible. $\omega = 1$ corresponds to fully reversible decoupling, where diffusion generates essentially one additional equivalent configuration and re-formation of the original vacancy–standing adjacency remains likely. $\omega \rightarrow 1$ indicates effectively irreversible decoupling, where diffusion delocalizes the vacancy such that re-establishing the original adjacency becomes negligible. An identical color scale is used for all panels to enable direct comparison. Boundaries between diffusion-limited (\textbf{DL}), reorientation-limited (\textbf{RL}), and adsorption-limited (\textbf{AL}) regimes are indicated by dotted lines, and the corresponding $\omega$ ranges are annotated. The thermodynamic phase diagram is shown in the background for reference, highlighting regions where standing (white) or lying (black) molecules are thermodynamically stable.}
  \label{fig:pub3_fig9}
\end{figure}

This enhanced decoupling is directly reflected in the effective values of $\omega$ obtained from the simulations. By inserting the simulated collective rate constants $\kLScol$ into the theoretical expression (Equation~\ref{eq:pub3_6}), $\omega$ can be extracted as an effective parameter that quantifies the (ir)reversibility of vacancy–molecule dissociation. The resulting values are shown in Figure~\ref{fig:pub3_fig9}.

For $f=3{:}1$, $\omega$ decreases with increasing temperature and decreasing pressure, ranging from about $1.5$ down to $0.007$. For $f=4{:}1$, even smaller values are obtained, spanning roughly $0.6$ to $0.004$ over the explored parameter space. 

The systematic decrease of $\omega$ at higher temperatures reflects the increasing rate constant of lying diffusion, which enhances vacancy propagation and renders vacancy–standing decoupling effectively irreversible. Lower pressures further promote this effect by reducing the rate of vacancy consumption via adsorption, thereby increasing the relative importance of diffusion-mediated redistribution.

\FloatBarrier

Small $\omega$ values therefore indicate that once diffusion becomes active, vacancies rapidly explore configurations that are no longer locally correlated with the original standing molecule. In this regime, not only is the re-formation of the original vacancy–standing adjacency unlikely, but the probability of re-establishing any configuration in which a standing molecule is adjacent to the required $(f-1)$ contiguous vacancies becomes strongly reduced. As a consequence, the local geometric condition necessary for back-reorientation is rarely fulfilled. In this limit, diffusion-induced vacancy delocalization effectively suppresses the reverse process, and stabilization of standing molecules dominates the kinetics. Consequently, the geometric prefactor $\gamma$ increases markedly with footprint ratio in diffusion-enhanced regimes, consistent with the order-of-magnitude accelerations observed in Figure~\ref{fig:pub3_fig8}.

\section{Conclusion}

This work addresses the central challenge of establishing a quantitative adsorbate-to-kinetics relationship for collective lying–standing transitions at organic–inorganic interfaces.
To derive such relations, we combined  kinetic Monte Carlo simulations with a systematic coarse-graining strategy: Simulated coverage-time profiles were reduced via an effective-two-state approximation to effective kinetic parameters and mapped onto a mean-field-like formulation that retains the effective geometric factors while remaining explicitly dependent on temperature- and pressure-dependent single-molecule rate constants.

For the reference system tetracyanoethylene/Cu(111), we identify a small set of microscopic channels that controls the collective transition. The early-stage kinetics is captured by a local two-step reorientation mechanism (stand-up followed by adsorption-driven stabilization), while the emergent collective rate is strongly modified by steric constraints that act on the reverse step. In regimes where lying diffusion is slow, vacancy-standing adjacency persists and the kinetics follows the purely local approximation. Once lying diffusion becomes faster than adsorption, diffusion stabilizes newly formed standing molecules by vacancy--molecule decoupling, thereby suppressing back-reorientation; at higher standing coverage, neighbor-induced back-reorientation introduces a self-inhibitory contribution that reduces the effective acceleration.

Systematic geometric variation reveals that molecular geometry provides an intrinsic and powerful handle to engineer collective transition timescales. Increasing overall molecular size accelerates transitions nearly proportionally to the lying footprint area, reflecting that each elementary event converts a larger fraction of the surface. More importantly, increasing the footprint ratio between lying and standing configurations strongly amplifies steric stabilization: Reorientation creates $\nvac=f-1$ contiguous vacancies and standing molecules sample the vacancy region quasi-uniformly, reducing effective back-reorientation multiplicities ($\nSL \propto 2/f$). In diffusion-enhanced regimes, larger footprint ratios additionally accelerate vacancy redistribution by lying diffusion, captured by a strongly decreasing effective multiplicity factor $\omega$, which quantifies how efficiently diffusion destroys vacancy-standing correlations.

These findings are condensed into an explicit algebraic expression for the collective rate constant that separates microscopic kinetics from geometric proportionalities. Because the formulation depends only on single-molecule rate constants and well-defined geometric parameters, it establishes a direct bridge between adsorbate–property relations and monolayer-wide transition times.


An important implication of this work is that the phase transitin rate constant can be related directly to well-known structure-to-property relationships for single molecules. Extension of the $\pi$-conjugated backbone, for example, often lowers the barrier for back-reorientation due to enhanced stabilization of the upright configuration.\cite{arefi_design_2022} Within our framework, this reduction directly accelerates the transition through the explicit dependence on $\kSL$. At the same time, larger backbones typically increase both the footprint ratio. Especially for higher temperature, this considerably increases the rate. This works in concert with the inreased molecular size, which again makes the phase transition markedly faster. 

Those effects are potentially counteracted by the fact that increasing molecular size commonly leads to higher diffusion barriers as a consequence of the larger molecule–substrate contact area.\cite{rosei_properties_2003,ruiz_density-functional_2016} A higher barrier for lying diffusion shifts the boundary between diffusion-limited and diffusion-enhanced regimes toward higher temperatures. As a result, diffusion-induced stabilization sets in only at elevated temperatures, effectively slowing down the collective transition under otherwise identical growth conditions. 
In addition, the molecular mass $m$ generally scales approximately with molecular area $A$. Larger molecules therefore exhibit higher adsorption rates at fixed pressure, with the temperature-dependence being negligible. This introduces a counterbalancing effect: Enhanced adsorption tends to accelerate stabilization of reoriented molecules, whereas the concurrent increase in diffusion barriers delays the onset of diffusion-assisted acceleration. The collective rate constant thus reflects the interplay of these competing geometric and energetic trends.

Because the collective rate constant depends explicitly on these elementary quantities, monolayer-wide transition times can be estimated directly from known or computable molecular parameters. In this way, collective kinetics emerges as a predictable consequence of established microscopic structure–property relations rather than as an opaque emergent phenomenon requiring full microkinetic simulations for each individual system.



\section{Methods}
\label{sec:p3_methods}

The collective lying–standing transition is modeled using the lattice-based kinetic Monte-Carlo (kMC) framework \texttt{kmos3},\cite{hoffmann_kmos_2014} which uses the Variable Step Size Method.\cite{bortz_new_1975,jansen_introduction_2003,gillespie_general_1976} The system evolves through stochastic events comprising adsorption, desorption, reorientation between lying and standing positions, and diffusion in both orientations.

All simulations are initialized from a fully formed monolayer of lying molecules. This reflects experimentally relevant growth scenarios, where deposition commonly leads to a metastable flat-lying phase before collective reorientation sets in. The analysis is restricted to temperature-pressure conditions where dense structures of standing adsorbates are thermodynamically favored, such that a transition from lying to standing adsorbate orientations can occur.

Adsorbates are represented as quasi-two-dimensional objects on a square lattice. Lying and standing molecules occupy multiple lattice sites, giving rise to steric constraints and spatial correlations at finite coverage. Reorientation occurs via edge positions of the molecule, corresponding to pivoting at the molecule-surface contact line. Lattice constants vary for model systems with different footprint ratios and sizes. Detailed information regarding lattice constants, super cell sizes, geometric aspects of single-molecule transitions are provided in Section \ref{sec:p3_SI1} of the Supporting Information.

The kMC simulations require a time acceleration algorithm, as rate constants of single-molecule events can vary by multiple orders of magnitude. This time disparity problem is tackled as proposed by Dybeck \textit{et al.}.\cite{dybeck_generalized_2017,andersen_assessment_2017}

Adsorption from the gas phase is treated as a non-activated process. Adsorption follows the kinetic gas impingement rate  formulated via
\begin{equation}
  k_{\mathrm{ads}} = s \frac{p a}{\sqrt{2 \pi m k_{\mathrm{B}} T}},
  \label{eq:pub3_kads}
\end{equation}
where $p$ is the partial gas-phase pressure of the adsorbed molecules, $a$ the adsorption area, $m$ the molecular mass, $s$ the sticking coefficient (set to unity throughout this work), $k_\text{B}$ the Boltzmann constant and $T$ the temperature. Adsorption is permitted only if sufficient contiguous lattice sites are available to accommodate the molecule in the respective orientation. Furthermore, adsorption can populate multiple molecular orientations. Therefore, the total gas-phase impingement rate is distributed among lying and standing adsorption channels using fixed scaling factors. Specifically, half of the adsorption flux is assigned to lying adsorbates, while the remaining half is equally divided between the two standing orientations (horizontally and vertically), preserving the total adsorption rate (see SI of ref \cite{werkovits_kinetic_2024}).

All microscopic rates are determined by the underlying energetic landscape and external growth parameters. The single-molecule rate constants of activated processes (diffusion, reorientation, and desorption) follow Arrhenius-type expressions, reading

\begin{equation}
  k = A \cdot \exp{\left(-\frac{\Delta E}{k_{\mathrm{B}}T}\right)},
\end{equation}

with the attempt frequency $A$, activation energy $\Delta E$, Boltzmann constant $k_{\mathrm{B}}$, and the temperature $T$. Activation energies of the reference system TCNE/Cu(111) are taken from ref \cite{werkovits_toward_2022}, schematically displayed in \ref{fig:pub3_fig1}, and summarized subsequently: Lying molecules diffuse slowly with a barrier $\dELL = \SI{0.45}{\electronvolt}$, whereas standing molecules are more mobile with a barrier $\dESS = \SI{0.05}{\electronvolt}$. Reorientation from lying to standing (standup) occurs with a barrier $\dELS = \SI{0.58}{\electronvolt}$, while the reverse process (laydown) has a much smaller barrier of $\dESL = \SI{0.04}{\electronvolt}$. For the sake of simplicity, a uniform attempt frequency of is used for all on-surface processes that differ from refs \cite{werkovits_toward_2022,werkovits_kinetic_2024}. Only the attempt frequency for desorption is defined by the barrierless adsorption rate (Equation~\ref{eq:pub3_kads}) to guarantee detailed balance with the adsorption rate constant $\kads$. The activation energy for desorption amounts to the absolute value of the corresponding adsorption energies $\EadsL = \SI{-2.40}{\electronvolt}$ and $\EadsS = \SI{-1.86}{\electronvolt}$ for adsorbates in a lying and standing orientation, respectively. For enhanced comparability, all on-surface processes are assigned the same attempt frequency of $A = \SI{1e12}{\hertz}$, a value commonly used for activated surface processes and similar to the ones determines in our previous work \cite{werkovits_toward_2022}.

Intermolecular interactions beyond steric exclusion are neglected to maintain generality and computational tractability. The simulations therefore capture the nucleation stage of the lying-standing transition. Additional stabilization from explicit intermolecular interactions would primarily accelerate growth once standing domains form. Accordingly, the collective rate constants extracted here can be interpreted as nucleation rates in the absence of growing seeds.

During each kMC simulation, snapshots are recorded to determine the surface areas occupied by lying (L) and standing (S) adsorbates, as well as, empty (E) spaces. These areas $a_{\mathrm{i}}$, with $\mathrm{i} \in \{ \mathrm{L}, \mathrm{S}, \mathrm{E}\}$, are converted into coverage fractions $\theta_\mathrm{i}=a_\mathrm{i}/a_\mathrm{tot}$, with the total area of the simulation supercell $a_\mathrm{tot}$. For each parameter set, five statistically independent trajectories are averaged to obtain smooth coverage-time profiles. For some simulations at $(T,p)$ points in the adsorption-limited regime (AL) only one trajectory is computed due to computational cost.

In the generated dataset, the adsorbate size $a_{\mathrm{L}}$ and footprint ratio $\f$ are systematically varied starting from the reference model system. For each parameter combination $(a_{\mathrm{L}},\f)$, simulations are performed on a continuous grid of temperature-pressure points spanning the region where standing adsorbates are thermodynamically stable. The sampled $(T,p)$ points and resulting kinetic regimes are documented in Section \ref{sec:p3_SI_kMCsampling} of the Supporting Information.

To obtain a compact and comparable description of the collective reorientation kinetics, the simulated coverage-time profiles are analyzed using an \textit{Irreversible Power-Law Two-State Approximation (IPL2SA)}. This approach is closely related to mean-field rate-equation descriptions \cite{dahl_surface_2000, felsen_model-free_2022} widely used in surface chemistry and heterogeneous catalysis, in which complex surface dynamics are mapped onto effective transitions between discrete states. Here, the full surface dynamics are coarse-grained into an effective transition from lying to standing orientations, capturing the essential kinetics while averaging over spatial correlations and steric constraints, which enter implicitly through effective kinetic parameters. The compactness and robustness of the IPL2SA formulation rely on two well-justified simplifications that are consistent with the conditions of our simulations: (i) irreversibility of the collective reorientation process and (ii) a two-state approximation. Irreversibility is justified because the local two-step process of reorientation is quasi-irreversible under the simulated conditions: Desorption is orders of magnitude slower than adsorption and on-surface rearrangements, and once a standing molecule is stabilized by adsorption, reversal to the lying orientation is sterically suppressed. Because the vicinity of the thermodynamic lying–standing phase boundary represents the most critical regime for the validity of the irreversibility assumption we added a discussion in \ref{sec:p3_SI_IPL2SA} of the Supplementary Information. The two-state approximation follows from neglecting empty surface regions in the effective rate equation. Although they are essential on the microscopic scale, their influence averages out in the effective description and they do not contribute quantitatively to the stabilization of collective reorientation events. Setting $\thetaE = 0$ reduces the surface balance from $\thetaS + \thetaL + \thetaE = 1$ to $\thetaL = 1-\thetaS$, yielding a two-state rather than a three-state description. This assumption is well satisfied in the simulations, where the maximum fraction of empty sites remains negligible ($\thetaEmax~=~0.03$). Altogether, this leads to the simple rate equation 

\begin{equation}
  \label{eq:pub3_2sa}
  \frac{\mathrm{d}\thetaS}{\mathrm{d}t} = \kLScol \thetaL^{\alpha},
\end{equation}

that describes the rate at which the coverage fraction of standing molecules $\thetaS$ changes in time $t$ as function of the coverage fraction of lying molecules $\thetaL$, the collective rate constant $\kLScol$ and apparent reaction order $\alpha$. An advantage of this approximation is that it admits an analytical solution,

\begin{equation}
  \theta_{\mathrm{S}}(t) =
  \begin{cases}
    1 - \exp\!\left(-\kLScol\, t\right)
    & \text{for }\alpha = 1 \\[6pt]
    1 - \left[(\alpha - 1)\, \kLScol \, t + 1\right]^{\tfrac{1}{1-\alpha}}
    & \text{for }\alpha \neq 1
  \end{cases}.
  \label{eq:pub3_thetaS}
\end{equation}

The collective rate constant $\kLScol$ integrates the combined influence of microscopic reorientation, adsorption, and spatial constraints into a single apparent kinetic parameter. The reaction order $\alpha$ quantifies how sensitively the transition rate depends on the coverage fraction of lying molecules. Values of $\alpha < 1$ or $\alpha > 1$ reflect cooperative or inhibitory collective effects, respectively, arising from steric constraints, vacancy formation, and spatial correlations on the surface. 

The analytical solution from Equation~\ref{eq:pub3_kads} is fitted to the averaged coverage-time data to obtain $\kLScol$ and $\alpha$. Fit quality is assessed using the coefficient of determination $R^2$, evaluated for coverage data that is quasi-equally spaced in time domain and visualized in SI Section~\ref{sec:p3_SI_IPL2SA}.

\section*{Acknowledgements}

We thank Sebastian Matera (Fritz Haber Institute, Freie Universität Berlin) for valuable and stimulating discussions on kinetic Monte Carlo simulations. Financial support from the Austrian Science Fund (FWF) through the START project Y1157-N36 is sincerely appreciated. Computational resources were provided by the Vienna Scientific Cluster (VSC), on which the kinetic Monte Carlo simulations were carried out.












  \clearpage
  \printbibliography[heading=bibintoc,title={References}]
\end{refsection}


\clearpage

\setcounter{page}{1}
\pagenumbering{arabic}

\begin{center}
  
  {\Huge \bfseries Supporting Information \par}
  \vspace{2.5em}
  {\LARGE \bfseries Emergent Rate Laws for Collective Lying–Standing Transitions \par}
  \vspace{1.2em}

  {\large
    Anna Werkovits\textsuperscript{1},
    Simon B. Hollweger\textsuperscript{1},
    Oliver T. Hofmann\textsuperscript{1,*}
    \par
  }
  \vspace{0.6em}
  {\itshape
    \textsuperscript{1}Institute of Solid State Physics, Graz University of Technology, 8010 Graz, Austria
    \par
  }
  \vspace{0.8em}
  {\small *Email: o.hofmann@tugraz.at \par}
\end{center}

\localtableofcontents

\begin{refsection}
  \setcounter{secnumdepth}{3}

  \renewcommand{\thesubsection}{S\arabic{subsection}}
  \renewcommand{\thesubsubsection}{\arabic{subsubsection}}

  \captionsetup[figure]{labelformat=suppl}
  \captionsetup[table]{labelformat=suppl}

\maketitle


\captionsetup[figure]{labelformat=suppl}
\captionsetup[table]{labelformat=suppl}
\setcounter{figure}{0}
\setcounter{table}{0}

\renewcommand{\thesubsection}{S\arabic{subsection}}
\renewcommand{\thesubsubsection}{\arabic{subsubsection}}


\section*{Supporting Information}
\label{sec:p3_SI}

\FloatBarrier

\subsection{Modeling details}
\label{sec:p3_SI1}

\begin{figure}[htb!]
  \centering
  \includegraphics[width=0.3\textwidth]{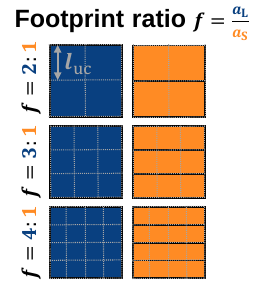}
  \caption{kMC representation of model systems with footprint ratios $f = 2$, 3, and 4. This figure extends Figure~\ref{fig:pub3_fig6} of the main text, where footprint areas of molecules in lying ($a_{\mathrm{L}}$) and standing ($a_{\mathrm{S}}$) orientations are denoted. For a footprint ratio $f = 2$, the lying molecule is represented by a $2 \times 2$ square of the kMC lattice (square lattice with lattice constant $l_{\mathrm{uc}}$), and by $3 \times 3$ and $4 \times 4$ squares for $f = 3$ and 4, respectively. Standing molecules occupy either $1 \times f$ or $f \times 1$ areas of the kMC lattice, corresponding to the two adsorption orientations of standing molecules (horizontal and vertical).}
  \label{fig:pub3_figS1}
\end{figure}

Figure~S\ref{fig:pub3_figS1} visualizes the kMC representation of model systems with different footprint ratios. The surface is modeled as a square lattice with lattice constant $l_{\mathrm{uc}}$, which defines the smallest unit of surface area accessible to the adsorbate. The lying--standing footprint ratio $f$ determines how many kMC lattice units are spanned by the molecular footprint. In the discrete kMC representation, this is implemented by mapping the footprint areas onto lattice units: Lying molecules occupy $f \times f$ contiguous lattice cells, while standing molecules occupy either $f \times 1$ or $1 \times f$ lattice cells, reflecting the two possible orientations of standing molecules on the surface. The footprint ratio thus directly determines how many lattice units are occupied by an adsorbate in each orientation and how vacancies generated during reorientation can be stabilized by adsorption. Further details on the kMC representation can be found in the Supporting Information of previous work.\cite{werkovits_kinetic_2024}

\begin{figure}[htb!]
  \centering
  \includegraphics[width=0.8\textwidth]{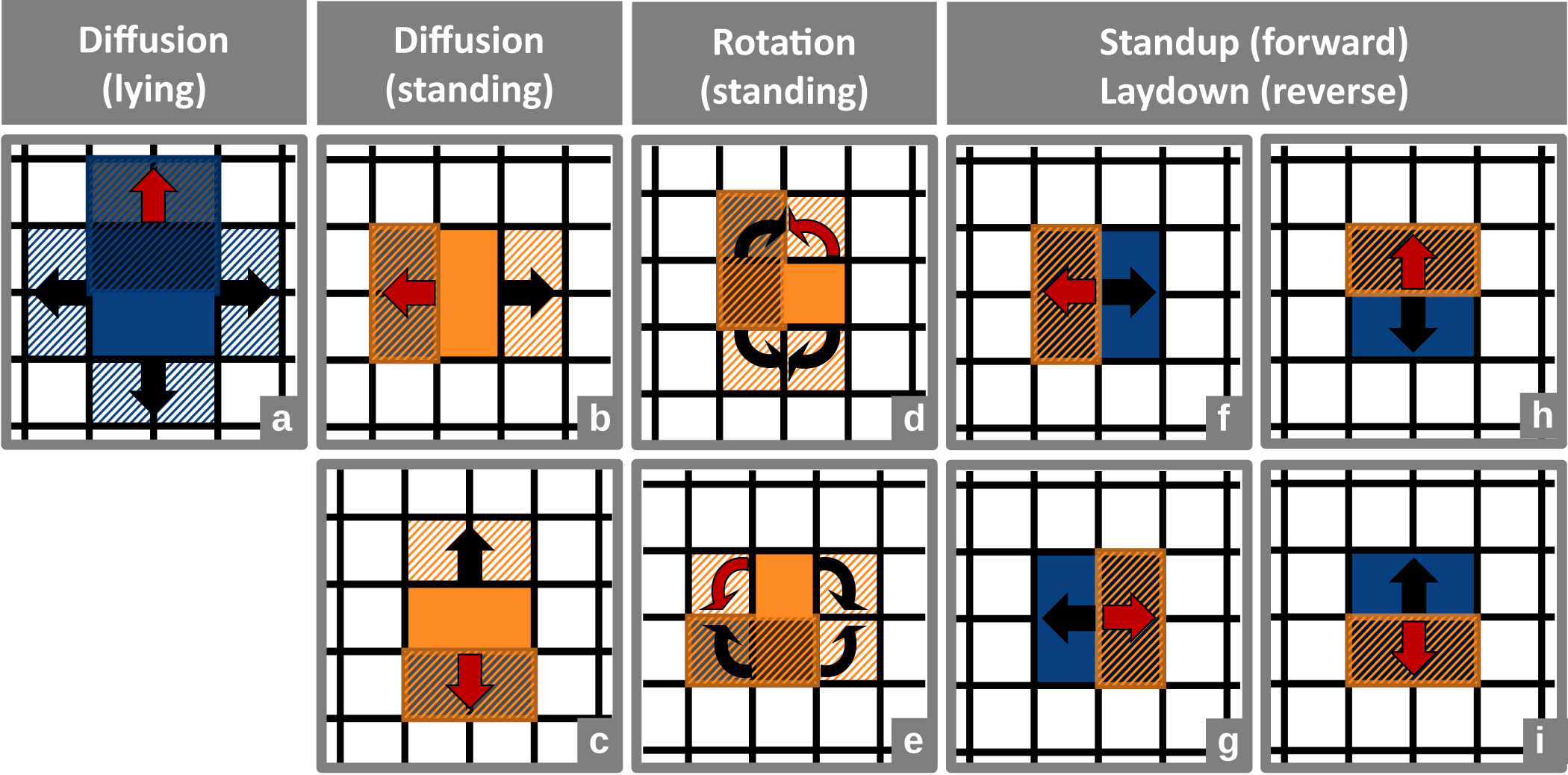}
  \caption{Definition of on-surface processes for the reference model system (footprint ratio $f = 2$) showing the three possible adsorption geometries lying, standing horizontal, and standing vertical. Shown are diffusion in lying (a) and standing (b-c) orientations, rotation in standing orientation (d-e), and reorientation from lying to standing (forward transition; f-i), referred to as stand-up, with the reverse transition termed lay-down. Initial adsorption sites are visualized as filled blue squares or orange rectangles, respectively, while final adsorption sites are shown with hatched shapes. Arrows indicate the transition directions. Only forward transitions are shown explicitly, whereas reverse transitions correspond to the inverted pathways. This schematic representation is adapted from the Supporting Information of ref~\cite{werkovits_kinetic_2024}.}
  \label{fig:pub3_figS2}
\end{figure}

Figure~S\ref{fig:pub3_figS2} illustrates the geometric definition of the on-surface processes for the reference model system (footprint ratio $f = 2$, molecular size $l_{\mathrm{L}} = \SI{6.8}{\angstrom}$, and lying adsorption energy $E_{\mathrm{ads,L}} = \SI{-2.4}{\electronvolt}$). Lying molecules are shown as blue squares, while standing molecules are shown as orange rectangles in horizontal or vertical orientation. Initial positions are indicated by filled shapes and final positions by hatched shapes, with arrows marking the possible transition directions. Model systems with larger footprint ratios behave analogously: Diffusion proceeds in steps of one kMC lattice unit, and reorientation always occurs via the edges.

The lattice constant of the reference system ($f_{\mathrm{ref}} = 2$) is $l_{\mathrm{uc,ref}} = \SI{3.4}{\angstrom}$, adopted from the coarse-grained kMC model of TCNE/Cu(111) introduced previously.\cite{werkovits_kinetic_2024} Note that this is an approximation arising from the use of a square lattice instead of the native hexagonal Cu(111) lattice, whose lattice constant is \SI{3.6}{\angstrom}. All simulations are designed such that, irrespective of the footprint ratio, $n_{\mathrm{L}}  \times n_{\mathrm{L}} = 12 \times 12 = 144$ molecules in the lying orientation fully cover the surface. The supercell size $n_{\mathrm{sc}}$ (number of repeating unit cells per dimension) is therefore given by
\begin{equation}
  n_{\mathrm{sc}} = n_{\mathrm{L}} \cdot f .
  \label{eq:pub3_S1}
\end{equation}

For footprint ratios $f = 2$, 3, and 4, this results in supercell sizes of $24 \times 24$, $36 \times 36$, and $48 \times 48$, respectively.
The lattice constants $l_{\mathrm{uc}}$ for larger footprint ratios $f = 3$ and 4 follow naturally Equation~\ref{eq:pub3_S2}. Therein, the lattice constant is obtained by dividing the size of the lying molecule $l_\mathrm{L}$ by the footprint ratio $f$.
This procedure yields lattice constants of \SI{2.27}{\angstrom} and \SI{1.70}{\angstrom} for footprint ratios of $f = 3$ and 4, respectively.

\begin{equation}
  l_{\mathrm{uc}} = \frac{l_{\mathrm{L}}}{f}
  \label{eq:pub3_S2}
\end{equation}

In the dataset where the molecular size $l_{\mathrm{L}}$ is varied, the lattice constant $l_{\mathrm{uc}}$ is adjusted according to Equation \ref{eq:pub3_S2}. The molecular size itself is varied in (half-)integer multiples of the lattice constant of the reference system, $l_{\mathrm{uc,ref}} = \SI{3.4}{\angstrom}$.
The resulting lattice constants are summarized in Table~S\ref{tab:pub3_tabS1}.

\begin{table}[!htbp]
  \centering
  \caption{Lattice constants $l_{\mathrm{uc}}$ for varying molecular sizes $l_{\mathrm{L}}$ and footprint ratios $f$.}
  \label{tab:pub3_tabS1}
  \sisetup{table-number-alignment=center}
  \begin{tabular}{c S S S}
    \toprule
    {$l_{\mathrm{L}}$ / \si{\angstrom}}
      & {$l_{\mathrm{uc}}$ ($f=2$) / \si{\angstrom}}
      & {$l_{\mathrm{uc}}$ ($f=3$) / \si{\angstrom}}
      & {$l_{\mathrm{uc}}$ ($f=4$) / \si{\angstrom}} \\
    \midrule
    $2.0 \times l_{\mathrm{uc,ref}}=6.8$  & 3.40 & 2.27 & 1.70 \\
    $2.5 \times l_{\mathrm{uc,ref}}=8.5$  & 4.25 & 2.83 & 2.13 \\
    $3.0 \times l_{\mathrm{uc,ref}}=10.2$ & 5.10 & 3.40 & 2.55 \\
    $3.5 \times l_{\mathrm{uc,ref}}=11.9$ & 5.95 & 3.97 & 2.98 \\
    $4.0 \times l_{\mathrm{uc,ref}}=13.6$ & 6.80 & 4.53 & 3.40 \\
    $4.5 \times l_{\mathrm{uc,ref}}=15.3$ & 7.65 & 5.10 & 3.82 \\
    \bottomrule
  \end{tabular}
\end{table}

To address the time-scale disparity arising from rate constants differing by many orders of magnitude, a time-acceleration algorithm is employed in the kMC simulations.\cite{dybeck_generalized_2017,andersen_assessment_2017} Because the number of standing molecules differs for models with different footprint ratios, the buffer parameter is adjusted to approximately the number of lattice sites in the corresponding simulation cell, rounded up to the next thousand. For the threshold, execution, and sampling parameters, default values were found to be sufficient. The explicit parameters are summarized in Table~S\ref{tab:pub3_tabS2}.

\begin{table}[htb!]
  \centering
  \caption{Time-acceleration parameters for kMC simulations with varying footprint ratios $f$.}
  \label{tab:pub3_tabS2}
  \begin{tabular}{l c c c}
    \toprule
     & {$f = 2$} & {$f = 3$} & {$f = 4$} \\
    \midrule
    \texttt{buffer\_parameter}     & 1000 & 2000 & 3000 \\
    \texttt{threshold\_parameter}  & 0.2  & 0.2  & 0.2  \\
    \texttt{execution\_parameter}  & 200  & 200  & 200  \\
    \texttt{sampling\_steps}       & 20   & 20   & 20   \\
    \bottomrule
  \end{tabular}
\end{table}

\newpage
\FloatBarrier

\subsection{kMC Sampling Details}
\label{sec:p3_SI_kMCsampling}
\FloatBarrier
\begin{figure}[htb!]
  \centering
  \includegraphics[width=1\textwidth]{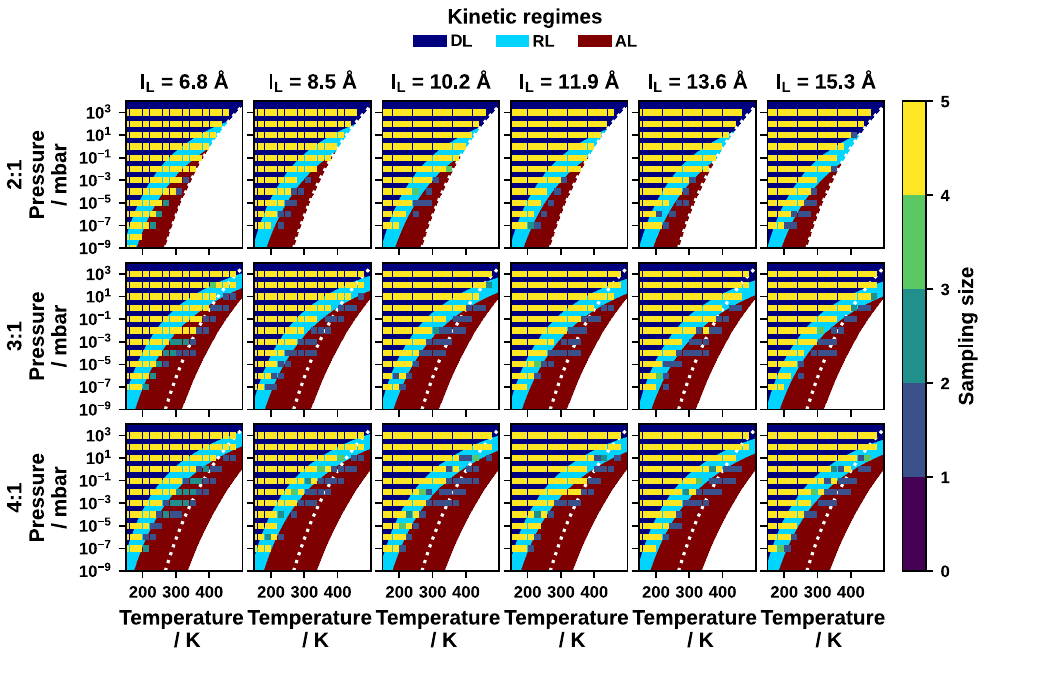}
  \caption{Kinetic Monte Carlo sampling size of temperature–pressure points.
Color coding indicates the number of independent simulations performed at identical temperature $T$, pressure $p$, adsorbate size $l_\mathrm{L}$ (columns), and footprint ratio $f$ (2:1, 3:1 and 4:1; rows). A target sampling size of five simulations was aimed for but is not reached in all cases due to computational cost. The background delineates the kinetic regimes (diffusion-limited, DL; reorientation-limited, light blue; adsorption-limited, dark red) and the thermodynamic stability region of the lying phase (white).}
  \label{fig:pub3_figS3}
\end{figure}
\FloatBarrier

\subsection{Irreversible Power-Law Two-State Approximation}
\label{sec:p3_SI_IPL2SA}


\subsubsection*{Validity near the thermodynamic lying–standing phase boundary}

As the irreversibility assumption is expected to be least accurate in the vicinity of the thermodynamic lying-standing phase boundary, we explicitly assess this region. In this regime, lying and standing domains coexist at long times due to phase equilibrium, and a finite backward flux is expected. Even under these conditions, the collective reorientation rate constants remain well captured by the IPL2SA framework. Deviations occur primarily in the extracted reaction order $\alpha$, reflecting the increasing influence of reversible fluctuations near coexistence. Within the irreversible two-state ansatz, this residual reversibility is effectively absorbed into the fitted reaction order, while the model drives the standing coverage toward $\thetaS \rightarrow 1$, although strictly $\thetaS < 1$ at coexistence due to phase balancing between lying and standing domains. In our analysis we are aware of these deviations, that are rather recognized as methodological artifacts than as indications of a breakdown of the effective description.

\newpage
\subsubsection*{Quality of IPL2SA Fits}
\glsreset{2sa}
\FloatBarrier

\begin{figure}[htb!]
  \centering
  \includegraphics[width=1\textwidth]{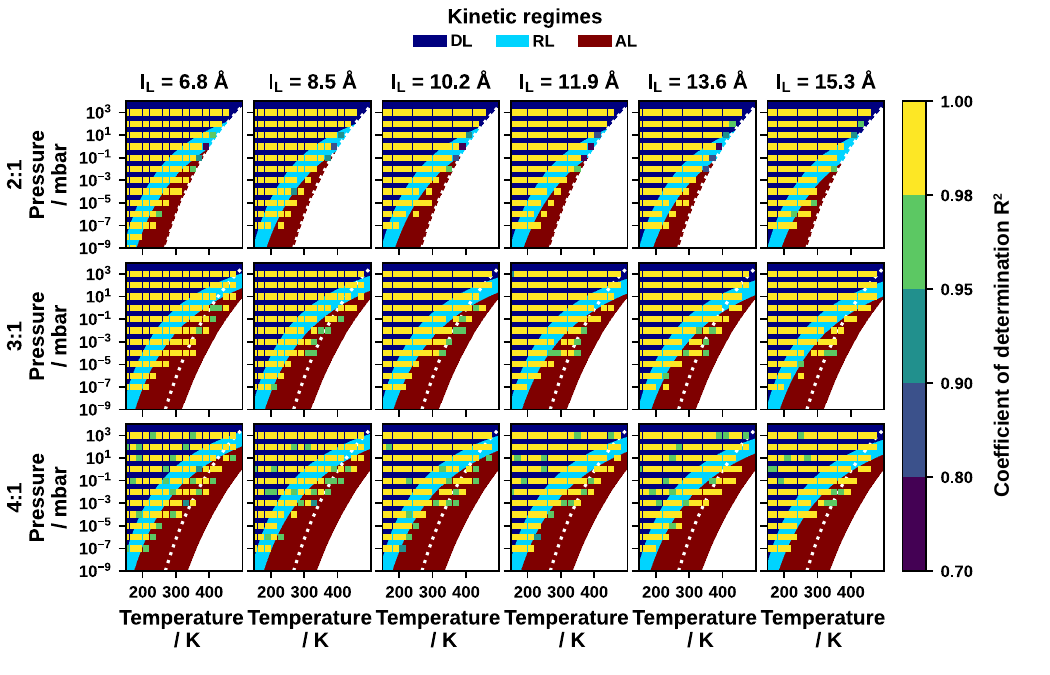}
  \caption{Quality of the \textit{\gls{2sa}} for the temperature–pressure points sampled by kinetic Monte Carlo.
The color scale quantifies the quality using the coefficient of determination $R^2$ for every combination of footprint ratios $f$ (2:1, 3:1 and 4:1) and adsorbate size $l_{\mathrm{L}}$. It is computed by comparing coverages from \gls{kmc} snapshots (approximately logarithmically spaced in time) with the corresponding \gls{2sa} predictions. The fit and $R^2$ evaluation are restricted to the most relevant regime of the lying–standing transition, namely the time interval where the standing coverage fraction $\thetaS$ lies within \numrange{0.5}{0.95}. The background delineates the kinetic regimes (diffusion-limited, dark blue; reorientation-limited, light blue; adsorption-limited, dark red) and the thermodynamic stability region of the lying phase (white).}
  \label{fig:pub3_figS4}
\end{figure}

\newpage
\FloatBarrier
\subsection{Apparent Reaction Orders}
\label{sec:p3_SI_alpha}
\FloatBarrier
\begin{figure}[htb!]
  \centering
  \includegraphics[width=1\textwidth]{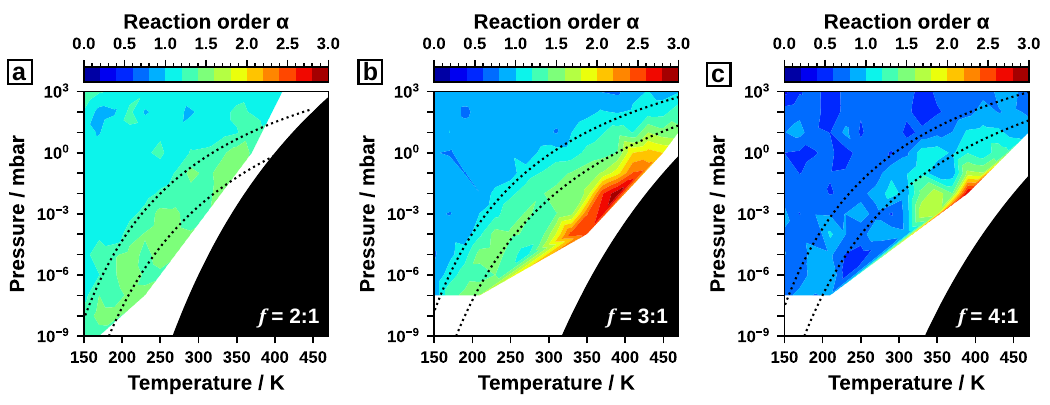}
  \caption{Pressure–temperature diagrams of the apparent reaction order $\alpha$ for model systems with increasing footprint ratios: (a) $\fpr{2}$, (b) $\fpr{3}$, and (c) $\fpr{4}$. $\alpha$ quantifies how strongly the transition depends on coverage and accordingly indicates steric effects like enhanced cooperativity ($\alpha < 1$) and inhibition ($\alpha > 1$). Note, that this is not a quantity from one microscopic process - rather it is an effective quantity from a collective transition obtained by the \textit{IPL2SA} via fitting (Equation~\ref{eq:pub3_2sa} and \ref{eq:pub3_thetaS}). An identical color scale is used for all panels to enable direct comparison. Boundaries between diffusion-limited (DL), reorientation-limited (RL), and adsorption-limited (AL) regimes are indicated by dotted lines, and the corresponding $\alpha$ ranges are annotated. The thermodynamic phase diagram is shown in the background for reference, highlighting regions where standing (white) or lying (black) molecules are thermodynamically stable.}
  \label{fig:pub3_figS5}
\end{figure}
\FloatBarrier

\newpage
\FloatBarrier
\subsection{Arrhenius Plots}
\label{sec:p3_SI_arrh}
\begin{figure}[htb!]
  \centering
  \includegraphics[width=1\textwidth]{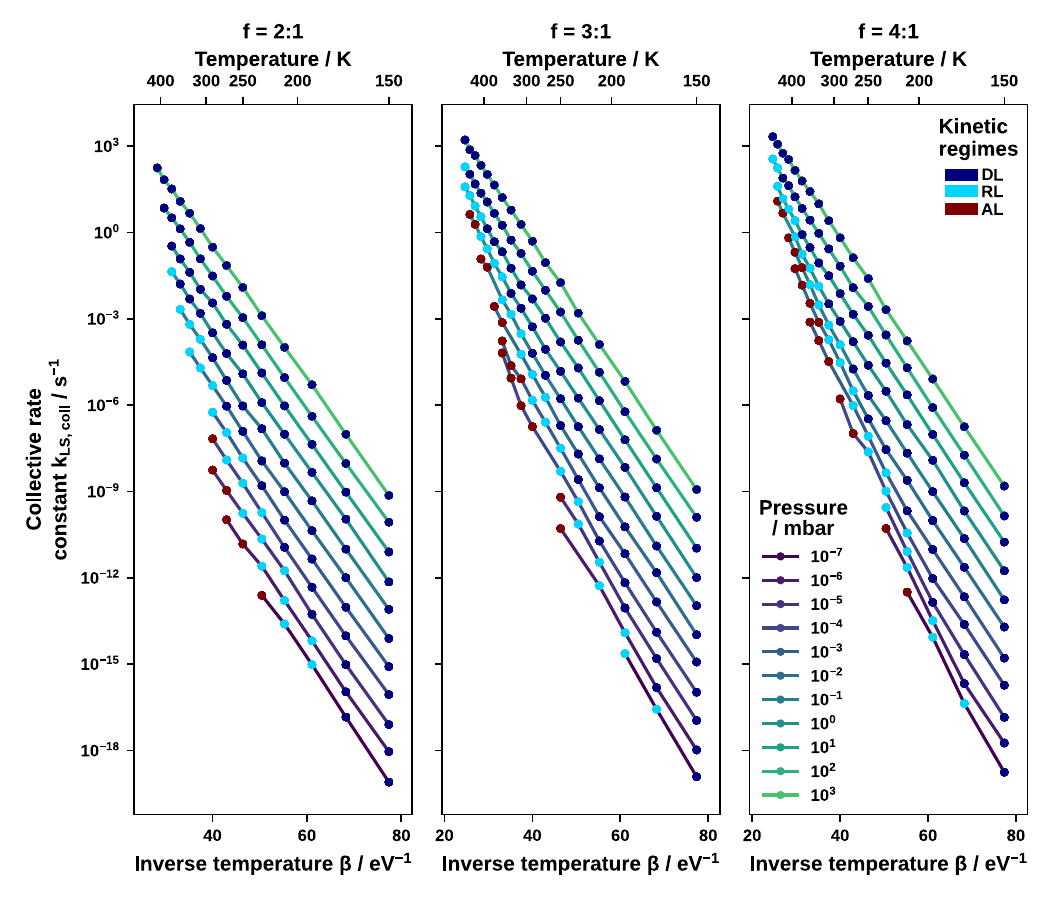}
  \caption{Arrhenius plots for footprint ratios $f = 2{:}1$, $3{:}1$, and $4{:}1$ at fixed adsorbate size $\l_{\mathrm{L}} = \SI{6.8}{\angstrom}$.
For each footprint ratio, collective rate constants are shown as a function of inverse temperature $\beta = (k_\mathrm{B} T)^{-1}$ (bottom axis) at different pressures (lines; see pressure legend). The corresponding temperature $T$ is indicated on the upper axis. Temperature–pressure points are assigned to their respective kinetic regimes, as indicated by the color coding (see legend).}
  \label{fig:pub3_figS6}
\end{figure}
\FloatBarrier
\newpage
\subsection{Derivation of effective diffusion-induced stabilization}
\label{sec:p3_SI_vacstab}

In the main text, diffusion of lying molecules is shown to stabilize newly formed standing molecules by removing the standing molecule from the vacancy created during reorientation (cf.\ States~$\ast_i$ in Figure~\ref{fig:pub3_fig5}). In this section, the corresponding correction to the local two-step reorientation model is derived in terms of an effective probability that a standing molecule can fall back into the lying configuration.

\subsubsection*{Probability for back-reorientation}
\label{subsec:p3_SI_pSL}

Back-reorientation (S$\rightarrow$L) requires that (i) a vacancy exists and (ii) the vacancy is adjacent to the standing molecule such that the lying footprint can be accommodated. We therefore write the probability that the back-reorientation condition is met as
\begin{equation}
  p_{\mathrm{SL}} = p_{\mathrm{vac}} \, \cdot p_{\mathrm{vac+S}},
  \label{eq:p3_SI_pSL}
\end{equation}
where $p_{\mathrm{vac}}$ denotes the probability that a vacancy exists, and $p_{\mathrm{vac+S}}$ the conditional probability that a vacancy is located next to a standing molecule (irrespective of the remaining local environment).

\subsubsection*{Complementary formulation via vacancy blocking by lying molecules}
\label{subsec:p3_SI_pvacS}

The probability $p_{\mathrm{vac+S}}$ can be expressed via its complementary event, namely that the site adjacent to the standing molecule is occupied by a lying molecule, which sterically blocks immediate back-reorientation. 
Defining $p_{\mathrm{vac+L}}$ as the probability that a lying molecule occupies the relevant neighboring site, one obtains
\begin{equation}
  p_{\mathrm{vac+S}} = 1 - p_{\mathrm{vac+L}}.
  \label{eq:p3_SI_pvacS_def}
\end{equation}

Within the effective description, $p_{\mathrm{vac+L}}$ is approximated by the ratio of the rate for establishing a blocking configuration (occupation by a lying molecule) and the total rate of all competing processes that prevent immediate back-reorientation. Concretely, the blocking event is governed by diffusion of lying molecules with rate constant $\kLL$. Competing processes comprise (i) adsorption of standing molecules into any of the $\nvac$ adsorption-enabled vacancies with rate $\nvac \kadsS$, and (ii) further lying-diffusion events that generate additional neighboring configurations, captured by an effective multiplicity factor $\omega$ multiplying $\kLL$. This yields
\begin{equation}
  p_{\mathrm{vac+L}}
  =
  \frac{\kLL}{\kLL + \nvac \kadsS + \omega \kLL},
  \label{eq:p3_SI_pvacL}
\end{equation}
and therefore
\begin{equation}
  p_{\mathrm{vac+S}}
  =
  1-\frac{\kLL}{\kLL + \nvac \kadsS + \omega \kLL}.
  \label{eq:p3_SI_pvacS}
\end{equation}


\subsubsection*{Effective back-reorientation propensity and geometric prefactor}
\label{subsec:p3_SI_gamma}

Using Eq.~\ref{eq:p3_SI_pSL}, the effective propensity for back-reorientation scales with $p_{\mathrm{SL}}$, such that the corresponding loss term can be written as
\begin{equation}
  \frac{\mathrm{d}p_1}{\mathrm{d}t}
  \propto
  \nSL \, \kSL \, p_{\mathrm{SL}}
  =
  \nSL \, \kSL \, p_{\mathrm{vac}} \, p_{\mathrm{vac+S}},
  \label{eq:p3_SI_kSLeff}
\end{equation}

Consequently, the diffusion-induced stabilization can be absorbed into the geometric prefactor $\gamma$ of the effective collective rate expression by renormalizing the probability for back-reorientation. In the notation of the main text this yields
\begin{equation}
  \gamma
  =
  \frac{\nvac \, \nLS}{\nSL \, p_{\mathrm{vac+S}}}
  \quad \text{with} \quad
  p_{\mathrm{vac+S}}
  =
  1-\frac{\kLL}{\kLL + \nvac \kadsS + \omega \kLL}.
  \label{eq:p3_SI_gamma_ext}
\end{equation}

Equations~\ref{eq:p3_SI_pvacS}--\ref{eq:p3_SI_gamma_ext} provide the diffusion-corrected contribution of vacancy--molecule decoupling to the effective geometric factor used to rationalize the regime-dependent collective kinetics.


  \clearpage
  \printbibliography[heading=bibintoc,title={References}]
\end{refsection}

\end{document}